# Performance Analysis of uniaxially strained Monolayer Black Phosphorus and Blue Phosphorus n-MOSFET and p-MOSFET

L. Banerjee*, A. Mukhopadhyay, A Sengupta, H Rahaman

Advanced Semiconductors & Computational Nanoelectronics Lab

School of VLSI Technology, IIEST

Shibpur, Howrah, West Bengal, India, Pin - 711103

*Corresponding author:  lopamudra.banerjee.rs2014@vlsi.iiests.ac.in

Abstract—

*In this work, we present a computational study on the possibility of strain engineering in monolayer Black Phosphorus (black P) and Blue Phosphorus (blue P) based MOSFETs. The material properties like band structure, carrier effective masses, carrier densities at band extrema are evaluated using Generalized Gradient Approximation (GGA) in Density Functional Theory (DFT). Thereafter self-consistent Non-Equilibrium Green's Function (NEGF) simulations are carried out to study the device performance metrics (such as output characteristics, ON currents, transconductance etc.) of such strained black P and blue P based MOSFETs. Our simulations show that carrier effective masses in blue P are more sensitive to strain applied in both zigzag and armchair directions. Blue P is more responsive in strain engineering for n-MOS and p-MOS. Except for black P based FETs with strain in armchair direction, overall the blue P (black P) n-MOSFET (p-MOSFET) show moderate to significant improvement in performance with tensile (compressive) strain in the transport directions.*

I. **INTRODUCTION**

Two-dimensional (2D) materials with atomic thickness such as monolayer Graphene, Phosphorene and transition metal dichalcogenides have aroused intensive research interests due to their fascinating electronic, mechanical, optical, or thermal properties, some of them not seen in their bulk counterparts. Because of atomic thickness, these materials also offer better electrostatic control than bulk materials, which make them interesting for fabrication of low-power electronic devices.

Due to the scaling limits of Si MOS technology, one of the critical and challenging issues in electronics industry is the development of alternative materials to improve the device performance. Phosphorene, the monolayer of Phosphorous, is promising for nanoelectronic applications not just because it is a natural p-type semiconductor but also because of its thickness dependent direct bandgap (in the range of 0.3 to 1.5 eV) [1]. The blue phosphorous (blue P), another single-layered allotrope of phosphorus, which was recently predicted to be nearly as stable as monolayer black phosphorous (black P). Predicted bandgap in blue P (~2eV) is also wider than that of the black P [2,3,4].

Tunability of electronic properties of 2D materials is very important for diverse applications. Externally induced and controlled strain can show interesting characteristics in terms of tuning/tailoring of the electronic properties as band structure, carrier effective masses, mobility etc [5]. Different approaches of inducing strain are lattice mismatch, functional wrapping [6,7], material doping [8, 9], and direct mechanical application [10]. Recently Liu et. al. [4] has predicted by ab-initio studies that a significant amount variation in the electronic properties of monolayer black P and blue P may be achieved with the application of strain. This prompts us to investigate how strain engineering may be employed to enhance the performance of monolayer black/ blue P FETs.

In the first part of this work, we compare through ab-initio studies how the material properties of black/ blue P are tuned under identical amount of strain [11,12]. We analyse the possible strain induced tuning of black/blue P material properties like band structure, carrier effective masses, carrier densities at band extrema using generalized gradient approximation (GGA) in Density Functional Theory (DFT). In the second part of the work using these computed parameters, transmission spectrum and output characteristics of n-MOSFET and p-MOSFET of black P and blue P were simulated. For device simulation studies we carry out self-consistent Poisson–Schrodinger solutions based on non-equilibrium Green's function (NEGF) formalism, as described [13,14].

## II. METHODOLOGY

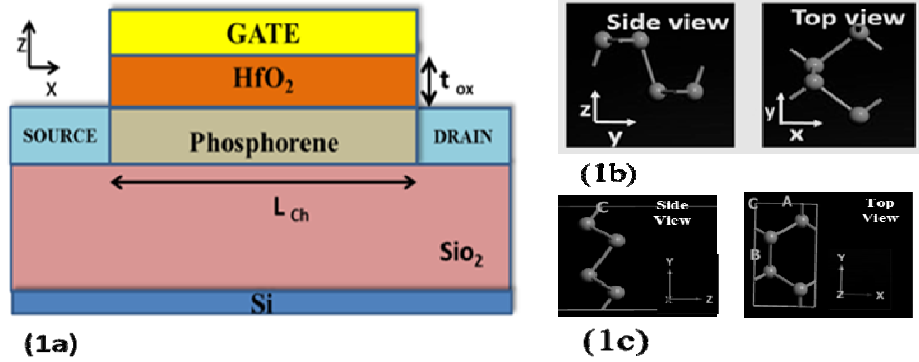

**Fig.1:** (a) The device schematic ($L_{Ch}$=15nm, $t_{ox}$ =2.5nm HfO$_2$, W=5nm, Doped source and drain).
(b) Monolayer black P unit-cell. (c) Monolayer blue P super cell.

For our study we consider a monolayer black P & blue P as channel material in MOSFET with channel length ($L_{Ch}$) of 15 nm (schematic shown in Fig. 1). The SiO$_2$/Si substrate is beneath the 2D channel. The gate dielectric of 2.5 nm thickness made of HfO$_2$ is over the channel material. We assume ideal doped contacts with aligned conduction band/valence band for the monolayer black P and blue P n-MOSFETs and p-MOSFETs [14].

**Material Simulation:**

For the material simulation, we use density functional theory (DFT) in Quantum Wise ATK [15]. We use a 9×9×1 Monkhorst-pack K-grid and Generalized Gradient Approximation (GGA) with Perdew-Bruke-Erzenhoff (PBE) exchange correlation function with a Double-Zeta Polarized (DZP) basis set. We set the grid-mesh cut off energy at 180 Rydberg and the Pulay mixer algorithm as iteration control parameter with tolerance value of $2\times10^{-4}$ Rydberg.

For the orthorhombic unit cell of black P the armchair direction is the y direction and the zigzag direction is the x direction (as in Fig. 1). As blue P has a hexagonal unit cell, thus to properly apply strain in the armchair/zigzag direction in hexagonal blue P, we construct a rectangular super cell of blue P containing four atoms. (Fig. 1).

We evaluate band structure and electron effective masses and other material properties of Phosphorus after applying varying tensile (+ve) and compressive (-ve) strain of 1%, 3%, 5%, 7% and 9% in the zigzag (A) and armchair (B) direction of the simple orthorhombic unit cell. Effective mass of electrons and holes are determined at three crystallographic directions <010>, <101> and <100> by parabolic fitting at the band extrema with the effective mass analyser in ATK. [15]

**NEGF Simulation:**

We obtain the effective masses of unstrained and strained black P and blue P from our DFT simulations, which are then used in our NEGF simulation studies to analyze the transistor performance of n-MOSFETs and p-MOSFETs. The simulation methodology is based on the framework proposed by Datta et. al. [13], Guo et. al. [14] and Ren et. al. [16]. The Green's function is written as

$$G(E) = [EI - H - \Sigma S - \Sigma D]^{-1} \qquad (1)$$

Here, I is the identity matrix, $\Sigma_{Source}$ and $\Sigma_{Drain}$ are self-energy matrices for the source and drain contacts and H is the effective mass Hamiltonian of the channel.

The transmission matrix, T(E) is represented as

$$T(E) = Trace[A_{Source}\Gamma_{Drain}] = Trace[A_{Drain}\Gamma_{Source}] \qquad (2)$$

Here, $\Gamma_{Drain}$ and $\Gamma_{Source}$ are the broadening matrices contributed by drain and source contacts respectively.

The Source (Drain) broadening,

$$\Gamma_{Source,Drain} = i(\Sigma_{Source,Drain} - \Sigma^{\dagger}_{Source,Drain})/2 \qquad (3)$$

$\Sigma^{\dagger}$ is Hermitean conjugate of $\Sigma$ matrix.

The Drain current [17],

$$I_D = \frac{e}{\hbar^2}\sqrt{\frac{m_y k_B T}{2\pi^2}} \int dE_{k_x} \left\{ F_{-1/2}\left(\frac{\mu_1 - E_{k_x}}{k_B T}\right) - F_{-1/2}\left(\frac{\mu_2 - E_{k_x}}{k_B T}\right) \right\} T_{SD}(E_{k_x}) \qquad (4)$$

Here, $h$ is the Planck's constant, $e$ is the electronic charge, $f_{Source}$ and $f_{Drain}$ are the Fermi functions in the source and drain contacts, $\eta_{Source}$ and $\eta_{Drain}$ are the source and drain chemical potentials, respectively. We consider fully ballistic transport for all our device simulation studies [18].

For device simulations, we consider both armchair and zigzag orientations of the channel (and therefore the transport direction). In the device the strain (tensile/ compressive) is assumed to be applied along the channel length (i.e. along transport direction only). This is due to the fact that a tensile (compressive) strain on the 2D phosphorous along the channel direction (say $x$) in a structure such as a MOSFET would be qualitatively equivalent to the application of compressive (tensile) strain perpendicular to the channel direction (i.e. $y$).

## III. RESULTS AND DISCUSSIONS

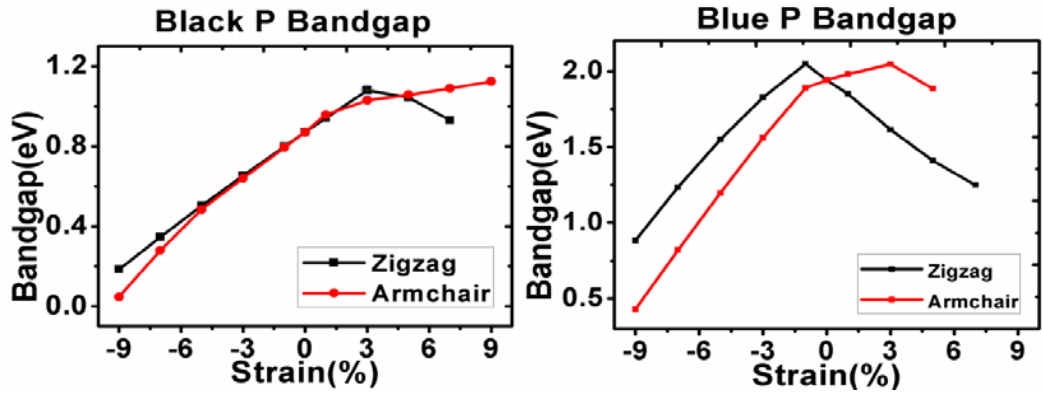

**Fig.2:** Bandgap variation in Black P and Blue P

In our simulation we observe the black P to be stable in the range of -9% to +7% strain for zigzag and –9% to +9% strain for armchair direction. Rectangular cell blue P are observed to be showed stable bonding integrity in the range of -9% to +7% stain for zigzag and -9% to +5% for armchair direction.

The band gap variations in black/blue P sheets due to application of strain is presented in Fig. 2 and in details (nature of band-gaps, band gap values) in Table I. In unstrained (0%) condition black P shows a direct band gap of 0.87 eV and blue P shows an indirect band gap of 1.94 eV (conduction band minima at $\Gamma$ point and valence band maxima at Y point of the rectangular supercell). These values are consistent with observations of Zhu et. al. [19] and Çakir et. al. [20].

For the range of strain under study, we observe for black P, compressive strain applied in zigzag direction, the band gap becomes indirect from -5% strain, while for tensile strain it remains direct in nature. The conduction band minima, remains at $\Gamma$ point throughout while the valence band maxima shifts between the $\Gamma$ point and the X point (in the orthorhombic cell) for strains $\geq$ -5%. When strain is applied in the armchair direction of black P, the band gap turns indirect for strains in excess of ±1%. Here the conduction band minima, remains at $\Gamma$ point throughout the range of applied strain, but the valence band maxima shifts to between the $\Gamma$ point and the X point for strains $\geq$ 3% and between -3% and -5 % . The valence band maxima however lies between $\Gamma$ point and the Y point for strains $\geq$ -7% in this case.

For blue P the nature of band gap remains indirect for the entire range of applied strain in the zigzag direction. The only difference being the valence band maxima which lies between the $\Gamma$ point and the X point (in the rectangular supercell) for compressive strains and shifts to between $\Gamma$ point and Y point for tensile strains. For strains in armchair direction in blue P, the band gap becomes direct (at $\Gamma$ point) for compressive strains, while it is indirect with valence band maxima between $\Gamma$ point and the Y point till 1% tensile strain. Over 1% applied strain in the armchair direction the valence band maxima shifts to points between $\Gamma$ point and the X point.

As for the values of the bandgaps for strain in the zigzag direction the maximum bandgap is obtained at +3% strain for black P (1.079 eV) and -1% strain for blue P (2.049 eV). For strain applied in armchair direction the maximum band gap (in the range of applied strain) for black P is observed at +9% strain (1.123 eV) for blue P this value is +3% strain (2.05 eV). The bandgaps in general show a tendency to close rather quickly with the application of compressive strain. Slight opening up of the gaps is observed for the tensile strain in armchair direction followed by a slight dip in bandgap, while for tensile strain in the zigzag direction, there exists the tendency of bandgaps to close (though at a much slower rate compared to that for compressive strain).

As referred in Table I and Figure 3, we observe that when strain applied in zigzag direction for Black P, bandgaps are indirect except at +3%. The effective mass of electron in (010) direction gradually increases from 0.04 to 0.18 for corresponding strain of -9% to +3%. However, at +5% effective mass suddenly increases to 1.21 and remains stable after that. For both (101) & (100) directions effective mass for electron remains stable in the range of 1.16 to 1.35 for strain between -9% to +3%. When subject to +5% strain, electron effective mass is decreased from 1.35 to 0.16 for (101) direction and from 1.22 to 0.14 for (100) direction. For black P effective mass for hole is exceptionally high i.e. 13.99 at 1% strain in zigzag in (101) and (100) directions. This value is ten times of that of effective mass for electron.

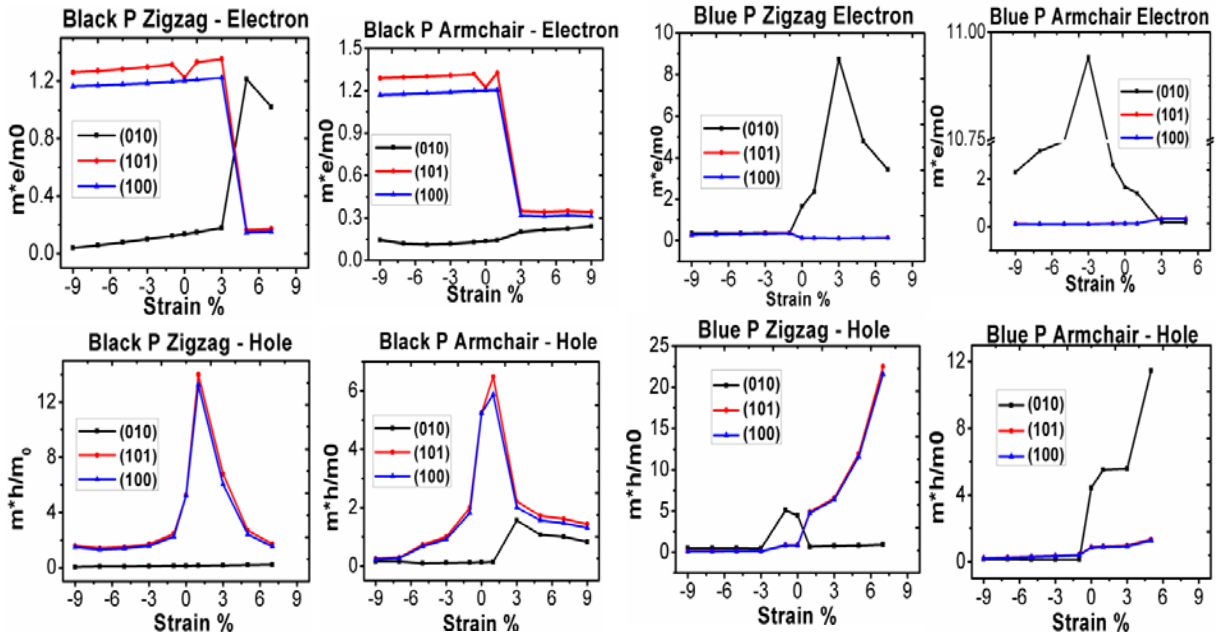

**Fig.3:** Effective mass variations for electrons & holes with strain in Black P & Blue P

For strain in Zigzag direction for blue P, bandgap remains indirect for strains ranging -9% to +7%. In (010) direction, effective mass of electron gradually increases from 0.36 to 2.37 against corresponding increase of strain from –9% to +1%. We observe a sudden increase of effective mass of electron at +3% to 8.75, which again on descending trend on further increase of strain. Other directions i.e. (101) & (100) electron effective mass of blue P are uniform in the range of 0.12-0.37. Effective mass of hole increases sharply for tensile strains in (101) and (100) directions. The maximum value of hole effective mass is 22.49 at 7% tensile strain in (101) direction. In (010) direction, hole effective mass is maximum (5.09) at -1% strain in zigzag direction.

For strain applied in armchair direction, for black P, bandgap remains Indirect for both tensile and compressive strains. In (010) direction, gradual increase (0.14 to 0.24) of electron effective mass is observed. In (101) direction effective mass is stable (1.28 to 1.32) between -9% and +1% strain. There is a sudden decreased from 1.32 to 0.34 at +3% & become stable after that. Effective mass variation in (100) direction is also very similar to that of (101), which experiences a sudden decrease from 1.20 to 0.31 at +3% strain. Hole Effective mass for black P, demonstrates much higher value (5.8 – 6.5) in (101) and (100) directions at +1% strain. Hole effective mass in (010) direction is similar to that of electron. In contrary, for blue P under strain in armchair direction, the bandgap is direct for compressive strains and changes to indirect when subject to tensile strain. In (010) direction, we observe a decreasing trend of effective mass from 2.28 to 0.16 against corresponding strain value of -9% to +5%, with an exception of sudden spike of 10.9 at -3% strain. We do not observe any sudden change of effective mass in (101) and (100) directions. For blue P for strain in armchair direction, hole effective mass in (010) direction is much higher (11.5) compared to (101) and (100) directions. The calculated effective masses are consistent with the findings of Zhu et. al. [19] and Çakir et. al. [20].

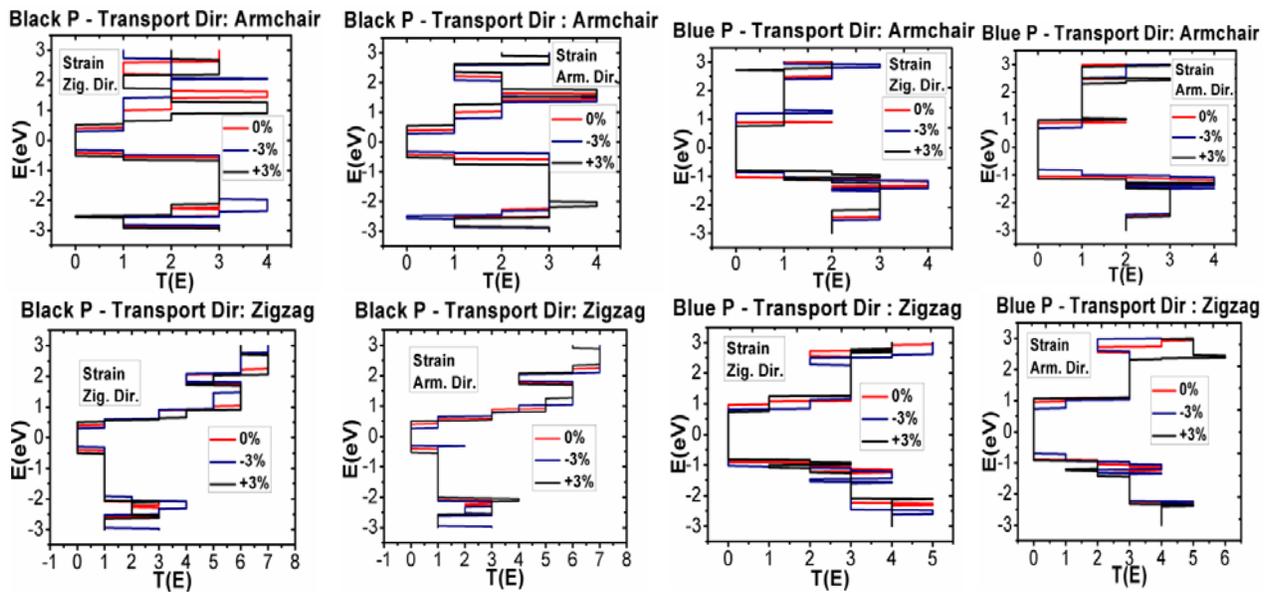

**Fig.4:** Transmission spectrum for Black & bBlue P in armchair & zigzag transport directions.

Impact of applied strains on transmission spectra has been illustrated in Fig 4. As the Fermi Level ($E_F$) is set as zero in our calculation the bias window is effectively between $-V/2$ to $+V/2$. In unstrained condition, in zigzag transport direction for black P it is observed that the area under transmission spectra in the corresponding bias window initially increases from that of zero bias value, but as the bias increased further, the same area seems to decrease with transmission mode becoming more sharply resolved along particular energy level, thus indicating possible drop in current density. In zigzag transport direction, in black P under tensile strain transmission modes are prominent near Fermi level, which indicates a greater transmission transparency of the channel. When subjected to compressive strain (both for the armchair and the zigzag direction), transmission modes move away from Fermi level thus indicating lesser conduction for low bias conditions. For blue P, though transmission modes are away from Fermi level, these have balanced distribution for electron and hole transport. In armchair transport direction, for black P, under tensile strain, high transmission modes are near Fermi level, thus showing greater transparency, whereas when subject to compressive strain, higher values are observed for electron transport. For blue P, same characteristics are observed for hole transport. We observe symmetric distribution of transmission spectrum for black P zigzag & blue P armchair transport direction. In contrary asymmetric transmission spectrum is observed for black P armchair and blue P zigzag transport direction.

  The symmetric or asymmetric behaviour of the transmission spectra can be attributed to the degree of match or mismatch of the electron and hole effective masses for the respective cases of applied strain type and direction and the direction of carrier transport.

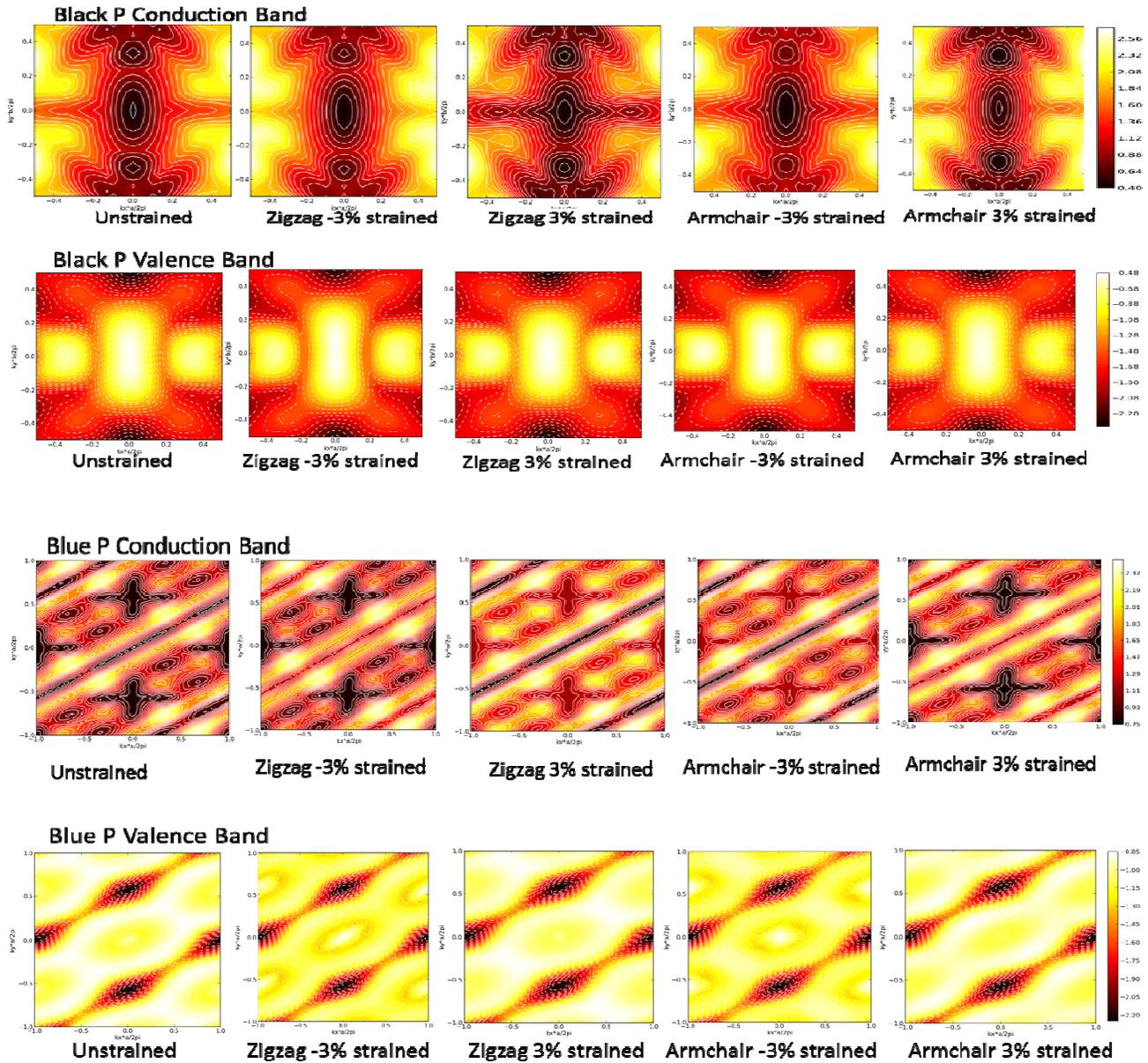

**Fig.5:** Carrier density plots in 1st Brillouin Zone

In Fig 5, diagrams show the surface plot of the valence and the conduction bands of blue and black P in the first Brillouin Zone (BZ) in unstrained as well as strained condition. From the plots we see a significant rising of the CB in blue P, with +3% strain in the zigzag direction and -3% strain in the armchair direction, uniformly through the first Brillouin zone. However for the -3% strain in zigzag and +3% strain in armchair there doesn't seem to be much overall change, but along the G-S path there exist a visible rise in the energy bands. From the plots we also see a significant rising of the CB in black P, with -3% strain in the zigzag direction and +3% strain in the armchair direction, uniformly through the first Brillouin zone. However for +3% strain in zigzag and -3% strain in armchair direction, CB moves downward with decreasing energies.

Decrease is noticeably greater with +3% in zigzag direction along the G-X path.

We observe a significant downward shift of the VB in blue P, with -3% strain in the zigzag and armchair direction, uniformly through the first Brillouin zone. However for the +3% strain in zigzag and armchair they seem to be much similar changes which is a prominent rise in the energy bands along the G-S path. Comparing both conduction and valence band, we see that in blue P over all energy of conduction band is much higher than valence bands. However of the VB for black P the ±3% strain in zigzag and ±3% strain in armchair there doesn't seem to be much overall change. Energy band share is rising in central position.

For device strain engineering in CMOS technology only small percentage of strain is generally applied to the channel for structural integrity of the device, hence we confine our studies on device simulation mostly to a moderate region of strain (-3% to +3%). This also ensures a stable configuration of the channel material itself as high strains i.e. 7- 9% are found to disintegrate bonds in our simulated structures.

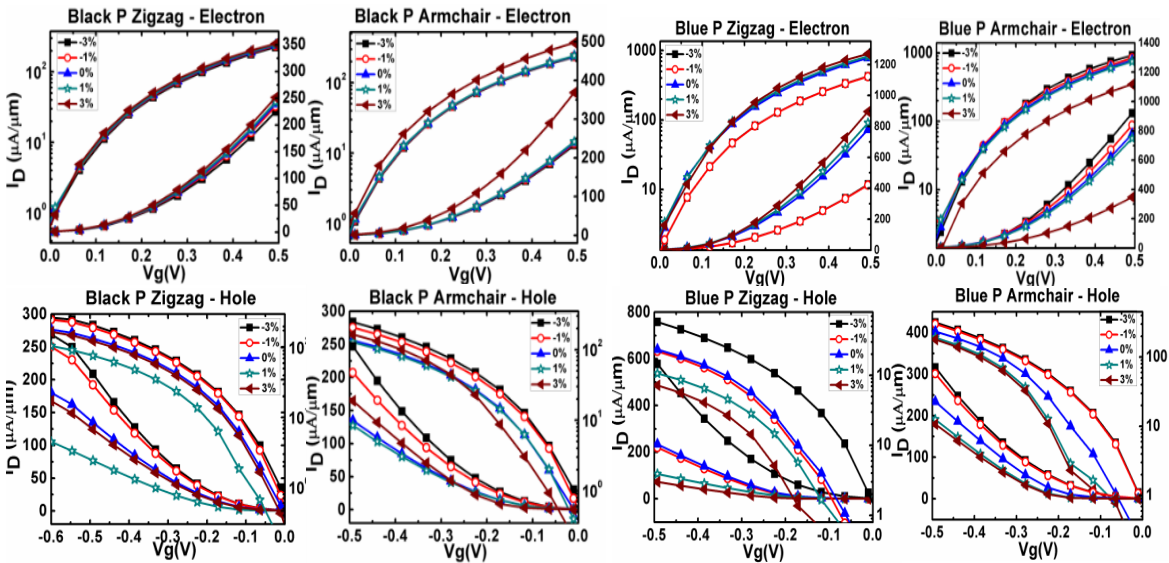

**Fig.6** Variation of drain current in linear scale and log scale with gate voltage in Black P and Blue P along Zigzag and Armchair directions.

$I_D$-$V_g$ output characteristics (in Fig 6.2) of n-MOSFET and p-MOSFET devices demonstrate the variation of the drain current for varying strain and gate voltage. We have simulated drain current under different uniaxial strains varying from -3% to +3%.

We observe that for black P in relaxed condition the drain current for n-MOSFET is 237 µA/µm and for p-MOSFET is 177 µA/µm. With the application of uniaxial strain of -3% in zigzag direction the drain current could be increased marginally for both n-MOSFET and p-MOSFET. For -3% strain in armchair direction drain current is almost doubled for both n-MOSFET and p-MOSFET to 500 µA/µm and 290 µA/µm respectively.

Blue P demonstrates much higher value of drain current for both n-MOSFET and p-MOSFET compared to black P. In relaxed condition in in zigzag direction the drain current is 779 µA/µm and 225 µA/µm for n-MOSFET and p-MOSFET respectively. At +3% strain in zigzag direction, the performance of n-MOSFET is marginally improved to 900µA/µm. For p-MOSFET exponential increase of drain current (600 µA/µm) is observed when subject to -3% strain which is 62% increase compared to relaxed condition. For armchair direction, drain current in unstrained condition are 800µA/µm and 240 µA/µm respectively for n-MOSFET and p-MOSFET. Our simulation reveals that under stained condition, for n-MOSFET drain current increases marginally to 930 µA/µm at -3% strain, whereas p-MOSFET shows 33% increase over relaxed condition.

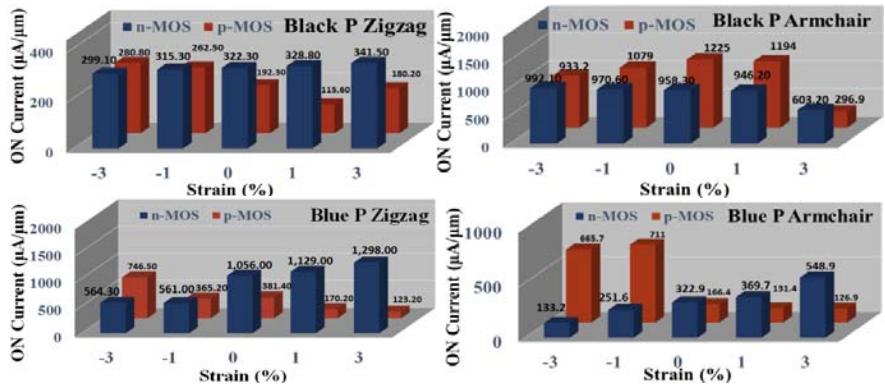

**Fig.7.** ON current variations with strain in Black P and Blue P along Armchair and Zigzag directions.

In Fig. 7, ON current variation is simulated against strains varying between -3% to +3% in both zigzag and armchair directions for both n-MOSFET and p-MOSFET. In relaxed condition for black P in zigzag direction ON current are 322.30 µA/µm and 192.30 µA/µm for n-MOSFET and p-MOSFET respectively. When simulated at +3% strain in zigzag direction n-MOSFET demonstrates a marginal increase of ON current to 341.50 µA/µm. For p-MOSFET ON current remains almost stagnant. In Armchair direction, ON current for Black P n and P MOSFETs are 3 times of that of zigzag direction, which are valued at 958.30 µA/µm and 1225 µA/µm respectively. When strain is applied in armchair direction, at +3% strain for n-MOSFET demonstrates 30% decrease on ON current to 603.20 µA/µm for n MOSFET and for p-MOSFET drop is much prominent of 80% with ON current of 296.9 µA/µm.

For blue P when simulated in zigzag strain, at relaxed condition, ON current are 1056 µA/µm and 381.40 µA/µm for n-MOSFET and p-MOSFET respectively. Under +3% strain in zigzag direction ON Current for n-MOSFET is increased marginally by 20% to 1298 µA/µm. When simulated at -3% strain p-MOSFET demonstrates considerable increase of 120% to 746.50 µA/µm. Strain simulation in armchair direction demonstrates similar trend as that of zigzag direction. ON current increases by 70% for n-

MOSFET +3% strain, whereas for p-MOSFET it shoots to 665.7 µA/µm (about 2 times increase) for strain variation to -3%.

Thus overall, black P shows slight improvement in n-MOS for zigzag direction tensile strain and similar improvement is observed for p-MOS for compressive strain in zigzag direction. Strain in armchair direction shows no significant improvement in both n and p-MOS.

For blue P in both zigzag and armchair transport directions, n-MOS shows improvement when subject to tensile strain and p-MOS under compressive strain. Specially in compressive strain in armchair direction blue P p-MOS demonstrates significant improvement.

The ON/OFF ratios for the various cases are calculated to be in the range ~ 0.6-1.8x10$^3$ for p-MOS and 1.2-1.7x10$^3$ for n-MOS devices. Our simulation reveals that for both n-MOSFET and p-MOSFET for black and blue P sub-threshold slope (SS) are not significantly impacted by externally induced uniaxial strains. In relaxed condition, both n-MOSFET and p-MOSFET are quite immune to short channel effects, with Drain Induced Barrier Lowering (DIBL) calculated to be within the range ~ 7.9 mV/V for p-MOS and ~8.25 mV/V for n-MOS. For these devices the sub-threshold slope (SS) was found to be in the range of ~ 61-67 mV/decade.

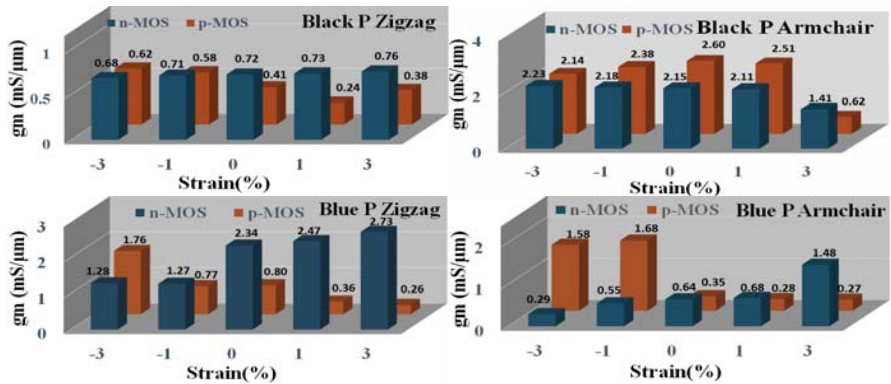

**Fig.8** Variation of transconductance (g$_m$) with gate voltage in Black P and Blue P electrons along Zigzag directions.

In Fig 8, we have shown the variation of simulated transconductance (g$_m$) with gate voltage (Vg=0.68 V) for n & p MOSFETs. For black P in zigzag direction the value of g$_m$ for n and p-MOSFET in relaxed condition are 0.72 mS/µm and 0.41 mS/µm respectively. For the strained condition at +3%, in zigzag direction, g$_m$ slightly increases to 0.76 mS/µm for n MOSFET, whereas for p MOSFET g$_m$ increases to 0.62 mS/µm when subject to -3% strain. Black P when strained in armchair direction, g$_m$ is higher for both n and p-MOSFET in relaxed as well as strained condition. In relaxed condition, g$_m$ for n and p-MOSFET are 2.25 mS/µm and 2.60 mS/µm respectively. When subject to +3% strain, gm for n-MOSFET drops by 35% to 1.41 mS/µm. But for p-MOSFET for similar strain reduction g$_m$ reduces sharply by 80% to 0.62 mS/um.

Variation of transconductance ($g_m$) in blue P exhibits very different characteristics for n and p MOSFETs when subject to strain in both zigzag and armchair direction. Transconductance ($g_m$) in relaxed condition are 2.34 mS/μm & 0.80 mS/μm for n and p MOSFET respectively in zigzag strain and 0.64 mS/μm & 0.35 mS/μm for armchair strain. At +3% strain in zigzag direction, for n MOSFET $g_m$ is increased marginally to 2.73 mS/μm, whereas for p MOSFET $g_m$ reduces to 0.26 mS/μm. On simulating at -3% transconductance for p-MOSFET is doubled to 1.76 mS/μm. For strain in armchair direction, at 3% compressive strain transconductance is more prominent for n MOSFET, which reaches to 1.48 mS/μm i.e. increase of 150%. Overall blue P n-MOSFET in zigzag transport direction and black P p-MOSFET in armchair transport direction show superior transconductance ($g_m$) values.

**Table I:** Effective Mass & bandgap of Black P and Blue P

| Black P: Strain in Zigzag | | Bandgap | Effective Mass - Electron | | | Effective Mass - Hole | | |
|---|---|---|---|---|---|---|---|---|
| Strain % | $CB_{min}/VB_{max}$ | Eg | (010) | (101) | (100) | (010) | (101) | (100) |
| -9 | Indirect (G-X) | 0.1843 | 0.0393 | 1.26045 | 1.16404 | 0.05562 | 1.59209 | 1.47002 |
| -7 | Indirect (G-X) | 0.3459 | 0.05806 | 1.27065 | 1.1695 | 0.07768 | 1.40306 | 1.29128 |
| -5 | Indirect (G-X) | 0.5028 | 0.07848 | 1.28281 | 1.17662 | 0.08609 | 1.5005 | 1.3772 |
| -3 | Direct (G) | 0.6524 | 0.10037 | 1.29708 | 1.18553 | 0.10196 | 1.70922 | 1.5658 |
| -1 | Direct (G) | 0.7987 | 0.1237 | 1.31352 | 1.19628 | 0.12042 | 2.45188 | 2.23947 |
| 0 | Direct (G) | 0.8702 | 0.13602 | 1.22528 | 1.20237 | 0.12893 | 5.23673 | 5.23425 |
| 1 | Direct (G) | 0.94 | 0.14878 | 1.3321 | 1.20879 | 0.14049 | 13.9887 | 13.2416 |
| 3 | Direct (G) | 1.079 | 0.17929 | 1.35244 | 1.22272 | 0.1645 | 6.79779 | 6.04858 |
| 5 | Direct (G) | 1.043 | 1.21373 | 0.16357 | 0.14738 | 0.18959 | 2.69383 | 2.41535 |
| 7 | Direct (G) | 0.9294 | 1.02043 | 0.17104 | 0.15353 | 0.21608 | 1.71165 | 1.53326 |

| Black P: Strain in Armchair | | Bandgap | Effective Mass – Electron | | | Effective Mass - Hole | | |
|---|---|---|---|---|---|---|---|---|
| Strain % | $CB_{min}/VB_{max}$ | Eg | (010) | (101) | (100) | (010) | (101) | (100) |
| -9 | Indirect (G-Y) | 0.04652 | 0.14462 | 1.2891 | 1.17191 | 0.15284 | 0.25146 | 0.22839 |
| -7 | Indirect (G-Y) | 0.2794 | 0.11835 | 1.29507 | 1.17733 | 0.16057 | 0.29392 | 0.267 |
| -5 | Indirect (G-X) | 0.484 | 0.11178 | 1.30199 | 1.18363 | 0.0933 | 0.7314 | 0.66722 |
| -3 | Indirect (G-X) | 0.6398 | 0.11734 | 1.30967 | 1.19061 | 0.11131 | 0.99871 | 0.91022 |
| -1 | Direct (G) | 0.7939 | 0.12901 | 1.31808 | 1.19826 | 0.12401 | 1.99562 | 1.81962 |
| 0 | Direct (G) | 0.8702 | 0.13602 | 1.22528 | 1.20237 | 0.12893 | 5.23673 | 5.23425 |
| 1 | Direct (G) | 0.9566 | 0.14334 | 1.32719 | 1.20654 | 0.13746 | 6.48367 | 5.87638 |
| 3 | Indirect (G-X) | 1.029 | 0.20042 | 0.34953 | 0.31785 | 1.57052 | 2.21045 | 2.01066 |
| 5 | Indirect (G-X) | 1.057 | 0.21593 | 0.34244 | 0.31142 | 1.07601 | 1.72504 | 1.56901 |
| 7 | Indirect (G-X) | 1.089 | 0.22436 | 0.35052 | 0.31876 | 1.00406 | 1.61858 | 1.47217 |
| 9 | Indirect (G-X) | 1.123 | 0.24059 | 0.34201 | 0.31104 | 0.82252 | 1.43972 | 1.30945 |

| **Blue P: Strain in Zigzag** | | **Bandgap** | **Effective Mass – Electron** | | | **Effective Mass - Hole** | | |
|---|---|---|---|---|---|---|---|---|
| **Strain %** | **$CB_{min}/VB_{max}$** | **Eg** | **(010)** | **(101)** | **(100)** | **(010)** | **(101)** | **(100)** |
| -9 | Indirect (G- X) | 0.8837 | 0.36167 | 0.28672 | 0.27836 | 0.50167 | 0.11404 | 0.11081 |
| -7 | Indirect (G - X) | 1.237 | 0.35865 | 0.31099 | 0.30148 | 0.48598 | 0.11649 | 0.11304 |
| -5 | Indirect (G - X) | 1.549 | 0.35732 | 0.33564 | 0.32489 | 0.47124 | 0.11824 | 0.11458 |
| -3 | Indirect (G - X) | 1.826 | 0.36347 | 0.35162 | 0.33989 | 0.45741 | 0.11955 | 0.11568 |
| -1 | Indirect (G - X) | 2.049 | 0.37019 | 0.36897 | 0.35615 | 5.09117 | 0.86055 | 0.83049 |
| 0 | Indirect (G - Y) | 1.941 | 1.65633 | 0.12923 | 0.1248 | 4.43098 | 0.87474 | 0.8436 |
| 1 | Indirect (G - Y) | 1.848 | 2.37066 | 0.13023 | 0.12565 | 0.68011 | 4.88465 | 4.72363 |
| 3 | Indirect (G - Y) | 1.611 | 8.7557 | 0.12062 | 0.11615 | 0.76902 | 6.56161 | 6.34287 |
| 5 | Indirect (G - Y) | 1.401 | 4.79259 | 0.12875 | 0.12378 | 0.80579 | 11.9422 | 11.546 |
| 7 | Indirect (G - Y) | 1.225 | 3.4226 | 0.13795 | 0.13241 | 0.94664 | 22.486 | 21.5636 |

| **Blue P: Strain in Armchair** | | **Bandgap** | **Effective Mass – Electron** | | | **Effective Mass - Hole** | | |
|---|---|---|---|---|---|---|---|---|
| **Strain %** | **$CB_{min}/VB_{max}$** | **Eg** | **(010)** | **(101)** | **(100)** | **(010)** | **(101)** | **(100)** |
| -9 | Direct (G) | 0.4313 | 2.28206 | 0.11441 | 0.11042 | 0.15119 | 0.21048 | 0.20311 |
| -7 | Direct (G) | 0.82 | 3.16764 | 0.11298 | 0.10904 | 0.14316 | 0.25195 | 0.24307 |
| -5 | Direct (G) | 1.198 | 4.94702 | 0.11191 | 0.10802 | 0.13531 | 0.30018 | 0.28955 |
| -3 | Direct (G) | 1.559 | 10.9413 | 0.11111 | 0.10725 | 0.12773 | 0.35578 | 0.34312 |
| -1 | Direct (G) | 1.888 | 2.58135 | 0.12697 | 0.12261 | 0.12004 | 0.41848 | 0.40353 |
| 0 | Indirect (G to Y) | 1.941 | 1.65633 | 0.12923 | 0.1248 | 4.43098 | 0.87474 | 0.8436 |
| 1 | Indirect (G to Y) | 1.98 | 1.40049 | 0.13084 | 0.12637 | 5.51189 | 0.92799 | 0.89491 |
| 3 | Indirect (G to X) | 2.05 | 0.16855 | 0.33988 | 0.32785 | 5.57566 | 0.96653 | 0.93207 |
| 5 | Indirect (G to X) | 1.875 | 0.16817 | 0.33962 | 0.32762 | 11.4536 | 1.32997 | 1.28227 |

## IV. CONCLUSION

We have studied the effect of varying tensile/compressive strain in both zigzag and armchair direction on black P and blue P n-MOS and p-MOS devices. Material properties are evaluated using DFT and the device characteristics are simulated through self-consistent Poisson-Schrodinger solution under NEGF formalism. Our studies show carrier effective masses in unstrained blue P is lower than that of black P thus giving better transport properties. Black P experiences an indirect to direct band gap transition at +3% strain in zigzag direction, whereas for blue P for strain above -1% in armchair crystallographic direction, experiences a direct to indirect bandgap transition. For black P effective mass for hole is more tunable than that of electron. For blue P, effective mass variation for both electron and hole are similar in both zigzag and armchair direction. We also observe high asymmetry (with suppressed hole transmission) in black P in the zigzag direction. In both zigzag and armchair transport directions in black P under tensile strain transmission modes become more prominent near Fermi level, which indicates a greater transmission transparency of the channel. Our simulation reveals that for both n-MOSFET and p-MOSFET for black and blue P sub-threshold slope (SS) are not significantly impacted by externally induced uniaxial strains. The n-MOS and p-MOS devices show good immunity to short channel effects (DIBL ~ 7.9-8.2 mV/V) and good sub-threshold slope (SS) in the range of ~ 61-67 mV/decade. The ON/OFF ratios for the various cases are calculated to be ~ 0.6-1.8x10$^3$ for p-MOS and ~1.2-1.7x10$^3$ for n-MOS devices. We predict an improvement of 20 % for n-MOS and 120 % for p-MOS in terms of ON currents at moderate strains of +3% and -3% depending on the transport direction. Improvements are also found in terms of enhancement of transconductance and ON/OFF ratios with the application of strain. The application of tensile strain in zigzag direction for blue P n-MOSFET is found to be most prominent for performance enhancement of the devices under study.

## V. ACKNOWLEDGMENT

The authors thank A. Jain & A.J.H. McGaughey of Department of Mechanical Engineering, Carnegie Mellon University for providing the structure of unstrained blue Phosphorous. This work was supported by DST under Grant No DST/INSPIRE/04/2013/000108 [IFA-13 ENG- 62].

**Figures caption:**

Fig.1. (a) The device schematic, (b) Monolayer black P unit-cell, (c) Monolayer blue P super cell.

Fig.2: Bandgap variation in Black P and Blue P

Fig.3: Effective mass variations for electrons &holes with strain in Black P & Blue P

Fig.4: Transmission spectrum for Black & bBlue P in armchair & zigzag transport directions.

Fig.5: Carrier density plots in 1st Brillouin Zone

Fig.6: Variation of drain current in linear scale and log scale with gate voltage in Black P and Blue P along Zigzag and Armchair directions.

Fig.7. ON current variations with strain in Black P and Blue P along Armchair and Zigzag directions.

Fig.8 Variation of transconductance (gm) with gate voltage in Black P and Blue P electrons along Zigzag directions.